\documentclass[journal]{IEEEtran}
\usepackage{etex}
\usepackage{tabularx} 
\usepackage{algorithm}
\usepackage{graphicx}
\usepackage[noend]{algpseudocode}
\usepackage{tikz}
\usetikzlibrary{shapes,arrows}
\usepackage{pstricks}
\usepackage{pst-node,pst-blur}
\usetikzlibrary{arrows}
\usepackage{amsfonts}
\usepackage{bm}
\usepackage{tabularx}
\usepackage{enumerate}
\usepackage{subfigure}
\usepackage{amsmath}
\usepackage{amssymb}
\usepackage{booktabs} 
\usepackage{multirow} 
\usepackage{mathtools} 
\usepackage{epsfig}
\usepackage{epstopdf}
\usepackage{array}

\usepackage[noadjust]{cite}
\usepackage{caption}
\usepackage{etoolbox}

\newtheorem{lemma}{Lemma}

\allowdisplaybreaks

\makeatletter 
\def\@eqnnum{{\normalsize \normalcolor (\theequation)}} 
\makeatother
\makeatletter
\def\set@curr@file#1{%
  \begingroup
    \escapechar\m@ne
    \xdef\@curr@file{\expandafter\string\csname #1\endcsname}%
  \endgroup
}
\def\quote@name#1{"\quote@@name#1\@gobble""}
\def\quote@@name#1"{#1\quote@@name}
\def\unquote@name#1{\quote@@name#1\@gobble"}
\makeatother
\usepackage{graphics}
\begin{document}

\title{Wireless Powered Protocol Exploiting Energy Harvesting During Cognitive Communications}
\author{Anirudh~Agarwal,~\IEEEmembership{Member,~IEEE}, and~Deepak~Mishra,~\IEEEmembership{Member,~IEEE}
\thanks{A. Agarwal is with the Department
of ECE, The LNM Institute of Information Technology, Jaipur, 302031 India (e-mail: anirudh.agarwal@lnmiit.ac.in}
\thanks{D. Mishra is with the Department of Electrical Engineering, Link\"{o}ping University, Link\"{o}ping, 58183 Sweden (e-mail: deepak.mishra@liu.se).}
}

\maketitle


\begin{abstract}
In this letter, a novel wireless powered protocol is proposed to maximize the system throughput of an energy harvesting (EH) based cognitive radio network, while satisfying a minimum primary user rate requirement. For EH, we exploit both dedicated wireless power transfer from primary base station as well as ambient ones available due to wireless information transfer among primary and secondary users. Specifically, we prove convexity of the optimization problem and obtain semi-closed-form for globally optimal solution. Numerical results validate the analysis, and show an average performance improvement of $70\%$ over benchmark scheme for various system parameters. 
\end{abstract}
\vspace{-1.5mm}
\IEEEpeerreviewmaketitle

\vspace{-5mm}
\section{Introduction}
\IEEEPARstart{A}{part} from spectrum scarcity, limited battery capacity of low-powered devices poses a major design challenge in 5G wireless communication systems. Cognitive radio (CR) is a promising solution to enhance spectrum utilization efficiency. Moreover, exploiting energy-harvesting (EH) techniques into CR networks (CRNs) has drawn wide attention due to ability of radio frequency (RF) signals to simultaneously carry energy as well as information. This enables better energy sustainability to both primary user (PU) and secondary user (SU) nodes, enhancing quality of service (QoS) in a CR network \cite{ccrn_tvt}. 

In \cite{6svm_tvt}, SUs first harvest energy from PU and then forward the primary data after amplification, while PU releases a portion of its time slot to SU in return. In \cite{8wcl}, wireless power transfer (WPT) from PU to a pair of SUs has been considered, and the problem of optimal time allocation (TA) has been solved while satisfying a constraint on outage probability of PU system. A joint power control and TA problem has been studied in \cite{wcl} for maximizing the SU throughput with a limiting interference constraint. However, one common limitation of \cite{6svm_tvt,8wcl,wcl} is that they have not considered a minimum average achievable rate requirement of PU. Shreshta et al. \cite{svm_tvt} did put a PU throughput constraint, but no closed-form expression was found for global optimal solution for TA, rather they obtained them numerically. Moreover, all the above works have either considered EH by SU from PU or EH by PU from SU, along with EH from primary base station (PBS).
\begin{figure}
     \centering
   \includegraphics[width=3.3in]{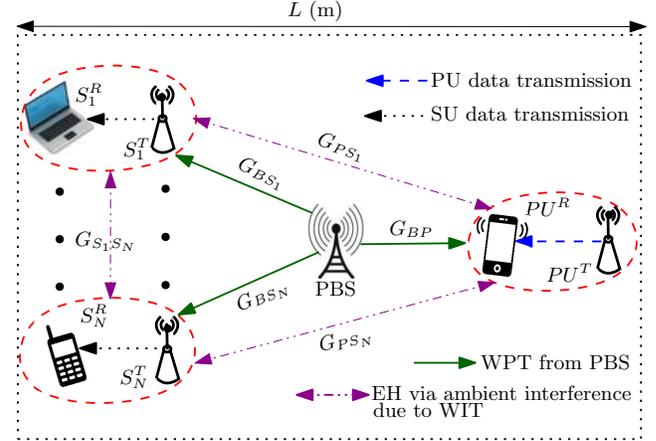}
     \caption{\small An EH-based CR system model with WPT and WIT.}
     \label{sysmod}
 \end{figure}
In this work, for better utilization efficiency of harvested energy, we have exploited all EH possibilities, i.e. EH via dedicated WPT from PBS and EH via ambient interference due to wireless information transfer (WIT) between every transmitter-receiver pair. This framework may be practically applicable to battery-constrained self-sustainable communication networks with CR \cite{adi_jsac}. Further, our optimal designs are targeted for serving applications with the overall system-centric goal, rather than individual node-level, where the best-effort delivery is desired to maximize the aggregate system throughput. \textit{To the best of our knowledge, this novel CR framework exploiting all RF-EH possibilities, along with minimum PU throughput constraint, has not been investigated earlier}. Key contributions of this work are: 1) A novel timing protocol is proposed with all RF-EH possibilities. 2) Convexity of optimization problem to maximize system throughput subject to a minimum PU rate constraint is proved. 3) Semi-closed-form globally optimal TA solution is obtained. 4) Improved performance of proposed scheme over benchmark and uniform TA schemes is numerically demonstrated with nontrivial optimal design insights.

\vspace{-2mm}
\section{Proposed Transmission Protocol}
We consider an EH-based interweave CR system with nodes scattered over a square field of length $L$ meters (m), as shown in Fig. \ref{sysmod}. It consists of a PBS at the center, a pair of primary transmitter $PU^{T}$ and receiver $PU^{R}$, and $N$ pairs of secondary transmitters $S^{T}_j$ and receivers $S^{R}_j$, $j\in \mathcal{N}_{N}$, where $\mathcal{N}_{N}\triangleq \{1, 2,...N\}$. PBS is assumed to be constantly powered by a source. Each PU and SU transmitter-receiver (TR) pair exhibits EH capabilities and is composed of a single omnidirectional antenna. The channel links between any two TR pairs are assumed to suffer from path-loss. Considering channel reciprocity, the channel power gains of the links from PBS to $PU^R$, PBS to $S^{R}_{j}$, $PU^T$ to $S^{R}_{j}$, and $S^{T}_{j}$ to $S^{R}_{k}$, are respectively denoted by $G_{BP}$, $G_{BS_{j}}$, $G_{PS_{j}}$ and $G_{S_{j}S_{k}}$, $ \forall j, k\in \mathcal{N}_{N}$. Here, we have assumed perfect channel knowledge for each communication link \cite{equalTA} with no external ambient EH source.
 \begin{figure}
     \centering
     \includegraphics[width=3.48in]{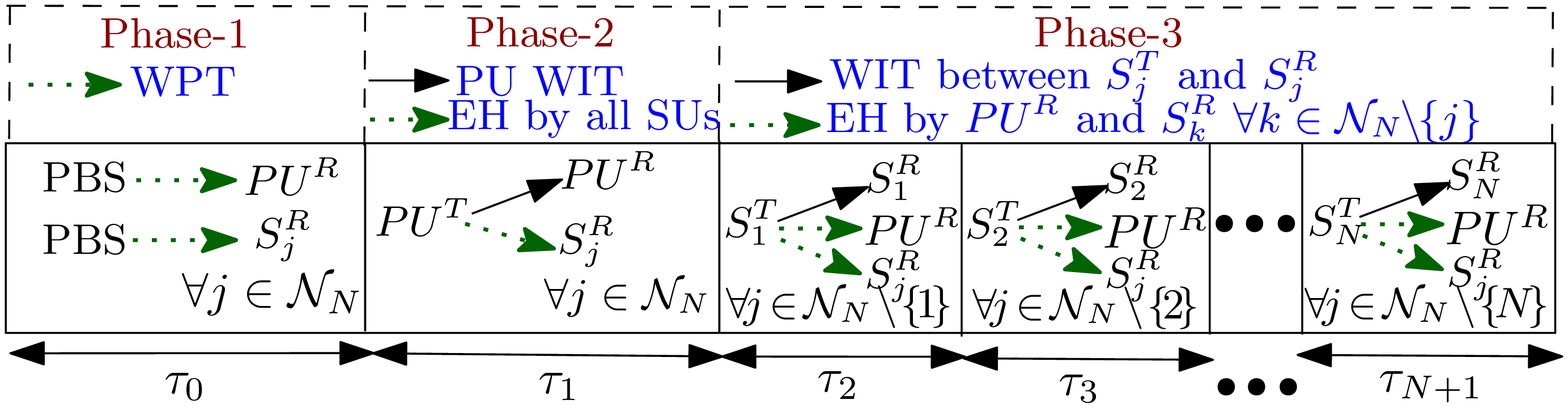}\vspace{-2mm}
     \caption{\small Proposed RF powered protocol for cognitive EH network.}
     \label{timediag}
 \end{figure}
In Fig. \ref{timediag}, the proposed novel timing protocol is depicted in which a 3-phase approach is followed for WPT and WIT. The time for WPT and WIT has been allocated orthogonally over a unit slot duration ($T=1$ sec). The optimization of TA for WPT and WIT is performed by the PBS on slot-by-slot basis. Further, we assume that each node is equipped with its own battery having sufficient stored energy in it, which if consumed, will be replenished later via ambient EH due to WIT in subsequent phases within a slot. So, the energy remains conserved in a slot. To be specific, in Phase-1, PBS transfers power wirelessly to PU and SUs for $\tau_{0}$ duration. In Phase-2 of $\tau_{1}$ duration, $PU^{T}$ can utilize its battery power for data transmission (DT), and SUs can harvest energy from ${PU}^T$-${PU}^R$ WIT. Similarly, every SU can consume its battery power for DT in their respective subphase of $\tau_{j}$ duration in Phase-3, $\forall j\in \{2,3,...N+1\}$. Whereas, PU can recharge its battery back during Phase-3 via EH due to the undergoing WIT between $S^{T}_j$ and $S^{R}_j$, $\forall j\in \mathcal{N}_{N}$. Likewise, every SU can transfer back its consumed power to its respective battery before the slots end, by EH during other nodes' DT phases. 
  
For WPT, with $P_{0}$ being the transmit power of PBS, energy harvested by PU from PBS over $\tau_{0}$ duration is given by,
\begin{equation}
  E^h_{BP}=\eta P_{0}G_{BP}\tau_{0}.
\end{equation}
where $\eta\in(0, 1)$ is the rectification efficiency.
Similarly, $E^h_{BS_{j}}\hspace{-0.5mm}\triangleq\hspace{-0.5mm} \eta P_{0}G_{BS_{j}}\hspace{-0.5mm}\tau_{0}$, $\forall j\hspace{-0.5mm}\in\hspace{-0.5mm}\mathcal{N}_{N}$, is the harvested energy by $j^{th}$ SU from PBS. Now, PU and SUs can further harvest ambient energy from each others WIT. Consequently, the power received by PU from $j^{th}$ SU, and that by $j^{th}$ SU from PU, can be respectively calculated as $P^R_{S_{j}P}\hspace{-0.5mm}=\hspace{-0.5mm}\frac{E^h_{BS_{j}}G_{PS_{j}}}{\tau_j}$, $P^R_{PS_{j}}\hspace{-0.5mm}=\hspace{-0.5mm}\frac{E^h_{BP}G_{PS_{j}}}{\tau_1}$. Moreover, $j^{th}$ SU receives $P^R_{S_{k}S_{j}}\hspace{-0.5mm}\triangleq \hspace{-0.5mm}\frac{E^h_{BS_{k}}G_{S_{k}S_{j}}}{\tau_k}$ power from the $k^{th}$ SU, $\forall j, k\hspace{-0.5mm}\in\hspace{-0.5mm}\mathcal{N}_N$, $k\hspace{-0.5mm}\neq\hspace{-0.5mm} j$. So, overall respective harvested energies for PU and SU DT are:
\begin{subequations}
\begin{gather}
E^h_{P}=E^h_{BP}+\sum_{j=1}^{N}P^R_{S_{j}P}\tau_{1},\label{eh_p}
\\
E^h_{S_{j}}=E^h_{BS_{j}}+P^R_{PS_{j}}\tau_j+\sum_{\substack{k\in\mathcal{N}_N}}^{k\neq j}P^R_{S_{k}S_{j}}\tau_j.\label{eh_s}
\end{gather}
\end{subequations}

Therefore, the PU throughput $R_{1}$ and SU throughput $R_{j+1}\forall j\in\mathcal{N}_{N}$ are defined as,
\begin{equation}\label{eq_rs}
\hspace{-2mm}R_{i}\hspace{-0.5mm}=\hspace{-0.5mm}\tau_{i}\hspace{-0.5mm}\left[\log_{2}\hspace{-0.5mm}\left(\hspace{-0.5mm}1\hspace{-0.5mm}+\hspace{-0.5mm}\frac{\gamma_{i}\tau_{0}}{\tau_{i}} \hspace{-0.5mm}\right )\hspace{-0.5mm} \right ]\hspace{-0.5mm},
\end{equation}
\noindent where $i\hspace{-0.5mm}\in\hspace{-0.5mm} \mathcal{N}_{1}\triangleq \mathcal{N}_{N} \cup \{N+1\}$, $\gamma_{1}\hspace{-0.5mm}=\hspace{-0.5mm}\frac{E^h_{P}G_{BP}}{\tau_{0}\,\sigma^{2}}$, $\gamma_{j+1}\hspace{-0.5mm}=\hspace{-0.5mm}\frac{E^h_{S_{j}}\,G_{S_{j}S_{j}}}{\tau_{0}\,\sigma^{2}}$ $\forall j\in\mathcal{N}_{N}$, and $\sigma^{2}$ is the variance of received zero mean additive white Gaussian noise. Now, from \eqref{eq_rs}, the system throughput $R_{sum}$ is defined as, $R_{sum}\hspace{-0.5mm}=\hspace{-0.5mm}\sum_{i=1}^{N+1}\hspace{-0.5mm}R_{i}$. Note that we have considered normalized bandwidth, so spectral efficiency has been defined in terms of throughput.

\section{Problem Definition}

\subsection{Mathematical Formulation}
To maximize $R_{sum}$ while guaranteeing a minimum PU rate requirement $\delta$, the following problem has to be solved:\\ \\
$\begin{aligned} 
\text{($\bm{\mathbb{P}}$)}\hspace{-1mm}:\hspace{1mm}&\underset{\bm{\tau}}{\text{max}} &&\hspace{-20mm}R_{sum},
&\hspace{-45mm}\text{subject to}
&&C1\hspace{-0.5mm}: R_{1}\ge \delta,\\& \hspace{0mm} C2\hspace{-0.5mm}:0\le \tau_{k} \le 1,&&\hspace{9mm}C3\hspace{-0.5mm}: \sum_{k=1}^{N+1}\tau_{k} \le 1,
\end{aligned}$\\
\noindent $\text{where}\hspace{0.5mm} \bm{\tau}\hspace{-0.5mm}=\hspace{-0.5mm}\left[\tau_{0}\, \,\tau_{1}\, \, \tau_{2}\, ...\, \tau_{N+1}\right]$. Keeping boundary constraint $C2$ implicit, the Lagrangian of ($\bm{\mathbb{P}}$) is given by, 

\vspace{-4mm}
\begin{eqnarray}
    &\hspace{-3mm}\mathcal{L}(\bm{\tau},\mu_{1},\mu_{2})=R_{sum}-\mu_{1}(\delta-R_{1})-\mu_{2}\left(\sum_{k=0}^{N+1}{\tau_{k}}-1\right)
\end{eqnarray}
\noindent where  $\mu_{1}$ and $\mu_{2}$ are the non-negative Lagrange multipliers associated with constraints $C1$ and $C3$ respectively. 

\subsection{Proof for Convexity of $\bm{(}\mathbb{P}\bm{)}$}
Here we provide Lemmas \ref{lem1} and \ref{lem2} for completing this proof.
\begin{lemma} \label{lem1} The objective $R_{sum}$ is a \textit{concave function} of $\bm{\tau}$. \end{lemma}
\begin{IEEEproof}
The Hessian $\mathbf{H}(Ri)$ of $R_{i}$ defined in \eqref{eq_rs}, $\forall i\in \mathcal{N}_{1}$ is a square matrix of order ($N+1$). So, the $H^{(i)}_{mn}$th element of  $\bm{H}\hspace{-0.5mm}\left(\hspace{-0.5mm}R_{i}\hspace{-0.5mm}\right)$ at $m^{th}$ row, $n^{th}$ column, $\forall m, n\in \mathcal{N}_{1}$ is given by:
\begin{eqnarray}\label{eq_a1}
H^{(i)}_{mn}\hspace{-0.5mm}=\hspace{-0.5mm}H^{(i)}_{nm}\hspace{-0.5mm}=\hspace{-0.5mm}\begin{cases}\hspace{-0.5mm}\frac{\partial^{2}\hspace{-0.5mm}R_{i}}{\partial\tau_{m}\partial\tau_{n}}\hspace{-0.5mm}=\hspace{-0.5mm}\gamma^2_{i}\tau_{0}\tau^{-2}_{i}\omega^{-2}_{i}\hspace{-0.5mm}; & \hspace{-1.5mm}\text{$\hspace{-0.5mm}m\hspace{-0.5mm}\neq n$, $\hspace{-0.5mm}m\hspace{-0.5mm}=\hspace{-0.5mm}i$, $\hspace{-0.5mm}n\hspace{-0.5mm}=\hspace{-0.5mm}1\hspace{-0.5mm}$},\\
    \hspace{-0.5mm}\frac{\partial^{2}\hspace{-0.5mm}R_{i}}{\partial\tau^2_{m}}\hspace{-0.5mm}=\hspace{-0.5mm}-\gamma^2_{i}\tau^{-1}_{i}\omega^{-2}_{i}\hspace{3.5mm}\,; & \hspace{-1.5mm}\text{$m\hspace{-0.5mm}=\hspace{-0.5mm}n\hspace{-0.5mm}=\hspace{-0.5mm}1$},\\
    \hspace{-0.5mm}\frac{\partial^{2}\hspace{-0.5mm}R_{i}}{\partial\tau^2_{m}}\hspace{-0.5mm}=\hspace{-0.5mm}-\gamma^2_{i}\tau^{2}_{0}\tau^{-3}_{i}\omega^{-2}_{i}\,; & \hspace{-1.5mm}\text{$m\hspace{-0.5mm}=\hspace{-0.5mm}n\hspace{-0.5mm}=\hspace{-0.5mm}i\ge 2$},\\
    \hspace{15mm}0\hspace{15mm}; & \hspace{-1.5mm}\text{otherwise},
    \end{cases}\hspace{-7mm}
\end{eqnarray}
\noindent where $\omega_{i}\triangleq1+\frac{\gamma_{i}\tau_{0}}{\tau_{i}}$. From \eqref{eq_a1}, it can be clearly seen that all the diagonal entries of $\bm{H}\hspace{-0.5mm}\left(\hspace{-0.5mm}R_{i}\hspace{-0.5mm}\right)$ are non-positive. Additionally, the determinants of all the odd principal minors of $\bm{H}\hspace{-0.5mm}\left(\hspace{-0.5mm}R_{i}\hspace{-0.5mm}\right)$ are non-positive and that of even principal minors are non-negative. Therefore, $\bm{H}\hspace{-0.5mm}\left(\hspace{-0.5mm}R_{i}\hspace{-0.5mm}\right)$ is a negative semi-definite matrix, and $R_{i}$ is a concave function of $\bm{\tau}$ \cite{boyd}. Moreover, as the sum of concave functions is also a concave function \cite{boyd}, $R_{sum}$ is thus proved to be a concave function of $\bm{\tau}$.
\end{IEEEproof}
\begin{lemma}\label{lem2} Constraints $C1$, $C2$, $C3$ are \textit{convex sets}. \end{lemma}
\begin{IEEEproof} The Hessian matrix of $R^{P}_{con}\triangleq\delta-R_{1}$, $\widehat{\bm{H}}\left(R^{P}_{con}\right)$ is given by:
\small $
\hspace{-0.5mm}\widehat{\bm{H}}\left(\hspace{-0.5mm}R^{P}_{con}\hspace{-0.5mm}\right)\hspace{-1mm}=\hspace{-1mm}\left[\hspace{-1mm} \begin{array}{ccc}
\hspace{-0.5mm}\frac{\hspace{-0.5mm}\partial^2 \hspace{-0.5mm}R^{P}_{con}\hspace{-0.5mm}}{\partial \tau_0^2} & \frac{\partial^2\hspace{-0.5mm} R^{P}_{con}}{\partial {\tau_0}\partial {\tau_1}} \\
\hspace{-0.5mm}\frac{\hspace{-0.5mm}\partial^2 \hspace{-0.5mm}R^{P}_{con}\hspace{-0.5mm}}{\partial{\tau_1}\partial {\tau_0}} & \frac{\hspace{-0.5mm}\partial^2\hspace{-0.5mm} R^{P}_{con}\hspace{-0.5mm}}{\partial \tau_1^2} \end{array} \hspace{-1.5mm}\right]\hspace{-1mm},$\normalsize \hspace{0.5mm}where $\frac{\hspace{-0.5mm}\partial^2\hspace{-0.5mm} R^{P}_{con}\hspace{-0.5mm}}{\partial \tau_0^2}\hspace{-1mm}=\hspace{-1mm}\frac{\hspace{-0.5mm}\gamma_1^2\hspace{-0.5mm}}{\tau_1\omega_1^2\hspace{-0.5mm}}$, $\frac{\partial^2 \hspace{-0.5mm}R^{P}_{con}}{\partial {\tau_0}\partial {\tau_1}}\hspace{-1mm}=\hspace{-1mm}\frac{\partial^2\hspace{-0.5mm} R^{P}_{con}}{\partial {\tau_1}\partial {\tau_0}}\hspace{-1mm}=\hspace{-1mm}\frac{-\gamma_1^2\tau_0}{\hspace{-0.5mm}\tau_1^2\omega_1^2}$, $\frac{\hspace{-0.5mm}\partial^2 \hspace{-0.5mm}R^{P}_{con}}{\hspace{-0.5mm}\partial {\tau_1}^2}\hspace{-1mm}=\hspace{-1mm}\frac{1}{\tau_1^3}\frac{\hspace{-0.5mm}\gamma_1^2\tau_0^2}{\hspace{-0.5mm}\omega_1^2}$.  So, it can be easily observed that $\frac{\hspace{-0.5mm}\partial^2\hspace{-0.5mm} R^{P}_{con}\hspace{-0.5mm}}{\partial \tau_0^2}, \frac{\hspace{-0.5mm}\partial^2\hspace{-0.5mm} R^{P}_{con}\hspace{-0.5mm}}{\partial \tau_1^2}\hspace{-1mm}>\hspace{-1mm}0$ and determinant of $\widehat{\bm{H}}\left(R^{P}_{con}\right)$ is zero. This proves that $\widehat{\bm{H}}\left(R^{P}_{con}\right)$ is positive semi-definite. Hence $R^{P}_{con}$ is a convex function of $\bm{\tau}$ \cite{boyd}. Further, constraints $C2, C3$ are linear in $\bm{\tau}$, therefore form convex sets. 
\end{IEEEproof}

\section{Globally-Optimal TA Solution}
As ($\bm{\mathbb{P}}$) is a convex optimization problem, the underlying Karush-Kuhn-Tucker (KKT) point ($\tau^*_{k},$ $\mu^*_{1},$ $\mu^*_{2}$) $\forall k\in \mathcal{N}_{0} \triangleq \mathcal{N}_{1} \cup \{0\}$, provides the globally-optimal solution of ($\bm{\mathbb{P}}$) \cite{boyd}. The corresponding KKT conditions are,
\begin{subequations}
\begin{gather}
    \frac{\partial \mathcal{L}}{\partial \tau_{k}}=0, \forall k\in \mathcal{N}_{0},\label{eq_kkt1}\\
    \mu_{1}(\delta-R_{1})=0,\label{eq_kkt2}\\
    \textstyle\mu_{2}\left(\sum_{k\in\mathcal{N}_0}{\tau_{k}}-1\right)=0.\label{eq_kkt3}
\end{gather}
\end{subequations}

Since the highest sum throughput can be obtained only by fully utilizing the available time resource, constraint $C3$ has to be satisfied at equality by the optimal solution, resulting into $\mu^*_{2} > 0$. Therefore from \eqref{eq_kkt3},
\begin{equation}\label{eq_obv}
    \textstyle\sum_{k=1}^{N+1}\;\tau_{k}=1-\tau_{0}.
\end{equation}
As $\mu^*_{i} \ge 0$, $\forall i\in\{1,2\}$, we next consider $2$ cases, i.e., unconstrained ($\mu^*_{1}=0$) and constrained ($\mu^*_{1}>0$) optimization problems with $R_{sum}$ as objective and $C1$ as constraint.

\vspace{-2mm}
\subsection{Unconstrained $R_{sum}$ Maximization}
For $\mu^*_{1}=0$, we solve \eqref{eq_rs} and \eqref{eq_kkt1} to get,
\vspace{-1mm}
\begin{eqnarray}\label{eq_sol1}
&\sum\limits_{j=1}^{N+1}\frac{\gamma_{j}}{1+\frac{\gamma_{j}\tau^*_{0}}{\tau^*_{j}}}=\mu^*_{2}\ln{2},
\end{eqnarray}
\vspace{-4mm}
\begin{eqnarray}\label{eq_sol2}
&\Phi\hspace{-0.5mm}\left(\hspace{-0.5mm}\frac{\gamma_{k}\tau^*_{0}}{\tau^*_{k}}\hspace{-0.5mm}\right)\hspace{-0.5mm}=\hspace{-0.5mm}\mu^*_{2}\ln{2},\,\,\,\forall \,\,k\in\mathcal{N}_1,
\end{eqnarray}
\noindent where $\Phi(x)\triangleq \ln{\left(\hspace{-0.5mm}1\hspace{-0.5mm}+\hspace{-0.5mm}x\right)}\hspace{-0.5mm}-\hspace{-0.5mm}\frac{x}{1+x}$. As $\Phi(x)$ is increasing in $x$, if $\Phi(x_{1})=\Phi(x_{2})$, then $x_{1}=x_{2}$, $\forall$ $x_{1}, x_{2}>0$. Thus, from \eqref{eq_sol2},
\begin{eqnarray}\label{eq_equal}
    &\frac{\gamma_{1}}{\tau^*_{1}}=\frac{\gamma_{2}}{\tau^*_{{2}}}=...=\frac{\gamma_{N+1}}{\tau^*_{(N+1)}}\triangleq K_{a}.
\end{eqnarray}
Now with $\Gamma_{a}\triangleq \sum_{k=1}^{N+1}\gamma_{k}$, from \eqref{eq_sol1}, \eqref{eq_sol2} and \eqref{eq_equal}, we have
\vspace{-0.5mm}
\begin{eqnarray}\label{eq_main}
   &\Phi\left(K_{a}\tau^*_{0}\right)=\frac{\Gamma_{a}}{1+K_{a}\tau^*_{0}},
\end{eqnarray}
\vspace{-0.5mm}
After few mathematical arrangements, \eqref{eq_main} reduces to,
\vspace{-1mm}
\begin{equation}\label{eq_k}
   1+K_{a}\tau^*_{0}= \frac{\Gamma_{a}-1}{\mathcal{W}\left(\frac{\Gamma_{a}-1}{\exp[1]}\right)} \triangleq f(\Gamma_{a}),
\end{equation}

\noindent where $\mathcal{W}(.)$ denotes the Lambert W function \cite{lambert}. So, from \eqref{eq_sol1} and \eqref{eq_k}, we obtain $\mu^*_{2}=\mu_{2a}\triangleq \frac{\Gamma_{a}}{(\ln{2})f\hspace{-0.5mm}\left(\Gamma_{a}\right)}$.

Additionally, from \eqref{eq_obv} and \eqref{eq_equal}, $K_{a}=\frac{\Gamma_{a}}{(1-\tau^*_{0})}$, which after substituting into \eqref{eq_k} will give,
\vspace{-2mm}
\begin{equation}\label{eq_tauE}
    \tau^*_{0}=\tau_{0a}\triangleq \frac{f(\Gamma_{a})-1}{\Gamma_{a}+f(\Gamma_{a})-1}.
\end{equation}
Next, \eqref{eq_equal} and \eqref{eq_tauE} can be solved to find $\tau^*_{k}=\tau_{ka}$ as below,
\vspace{-1.5mm}
\begin{equation}\label{eq_tauI}
    \tau_{ka}\triangleq \frac{\gamma_{k}}{{\Gamma_{a}+f(\Gamma_{a})-1}} ,\forall k\in\mathcal{N}_1.
\end{equation}
So, if $R_{1}\hspace{-0.5mm}>\hspace{-0.5mm}\delta$, then ($\tau^*_{k}\hspace{-0.5mm}=\hspace{-0.5mm}\tau_{ka},$ $\mu^*_{1}\hspace{-0.5mm}=\hspace{-0.5mm}0,$ $\mu^*_{2}\hspace{-0.5mm}=\hspace{-0.5mm}\mu_{2a}$) $\forall k\in \mathcal{N}_{0}$, is a feasible KKT point, and thus, the optimal TA solution of ($\mathbb{P}$).

\subsection{Primary Constrained $R_{sum}$ Maximization}
Here $C1$ is tight, which from \eqref{eq_rs} and \eqref{eq_kkt1} results in,
\vspace{-2mm}
\begin{eqnarray}\label{2eq_obv}
    &\tau^*_{0}=\frac{\tau^*_{1}}{\gamma_{1}}\left(2^{\frac{\delta}{\tau^*_{1}}}-1\right),
\end{eqnarray}
\vspace{-2mm}
\begin{eqnarray}\label{2eq_sol1}
&\sum\limits_{j=1}^{N+1}\frac{\gamma_{j}}{1+\frac{\gamma_{j}\tau^*_{0}}{\tau^*_{j}}}+\frac{\mu^*_{1}\gamma_{1}}{1+\frac{\gamma_{1}\tau^*_{0}}{\tau^*_{1}}}=\mu^*_{2}\ln{2},
\end{eqnarray}
\vspace{-2mm}
\begin{eqnarray}\label{2eq_sol2}
&\left(1+\mu^*_{1}\right)\Phi\hspace{-0.5mm}\left(\hspace{-0.5mm}\frac{\gamma_{1}\tau^*_{0}}{\tau^*_{1}}\hspace{-0.5mm}\right)\hspace{-0.5mm}=\hspace{-0.5mm}\mu^*_{2}\ln{2},
\end{eqnarray}
\begin{equation}\label{2eq_sol3}
\Phi\hspace{-0.5mm}\left(\hspace{-0.5mm}\frac{\gamma_{k}\tau^*_{0}}{\tau^*_{k}}\hspace{-0.5mm}\right)\hspace{-0.5mm}=\hspace{-0.5mm}\mu^*_{2}\ln{2},\,\,\,\forall k\in\{2,3,...N+1\}.
\end{equation}
Similar to \eqref{eq_equal}, equation \eqref{2eq_sol3} can be rewritten as,
\vspace{-1.5mm}
\begin{equation}\label{2eq_equal}
    \frac{\gamma_{2}}{\tau^*_{2}}=\frac{\gamma_{3}}{\tau^*_{{3}}}=...=\frac{\gamma_{N+1}}{\tau^*_{(N+1)}}\triangleq K_{b}.
\end{equation}
With $\Gamma_{b}\triangleq \sum_{k=2}^{N+1}\gamma_{k}$, from \eqref{eq_obv}, \eqref{2eq_sol1} and \eqref{2eq_equal}, we obtain:
\vspace{-0.5mm}
\begin{eqnarray}\label{2eq_main1}
    &\frac{\left(1+\mu^*_{1}\right)\gamma_{1}}{1+\frac{\gamma_{1}\tau^*_{0}}{\tau^*_{1}}}+\frac{\Gamma_{b}}{1+K_{b}\tau^*_{0}}=\mu^*_{2}\ln{2},
\end{eqnarray}
\begin{figure}[b!]
    \centering
    \includegraphics[width=3.48in]{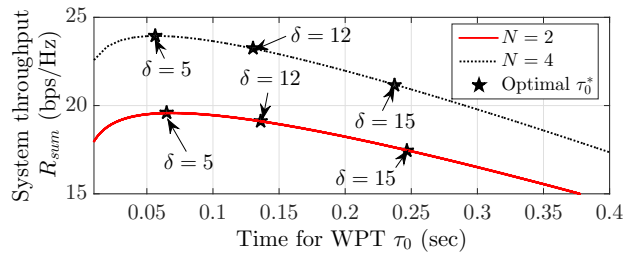}
    \caption{\small Variation of $R_{sum}$ with $\tau_0$, and impact of $\delta$ on optimal TA.}
    \label{valid}
\end{figure}
\begin{figure}
    \centering
    \includegraphics[width=3.1in]{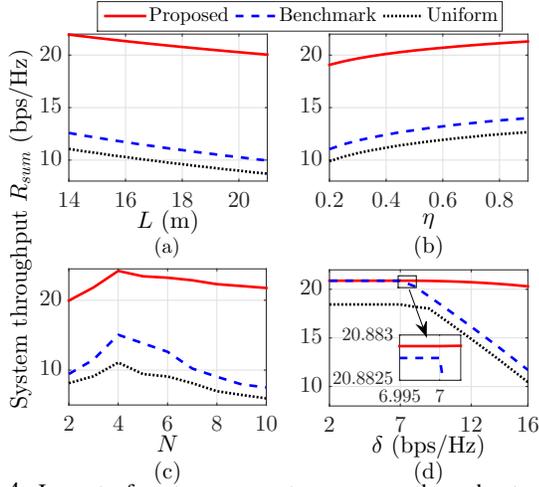}\vspace{-4mm}
    \caption{\small Impact of system parameters on sum throughput and comparative analysis of different TA schemes.}
   \label{impact}
\end{figure}
\hspace{-1.5mm}Further, \eqref{2eq_sol3} and \eqref{2eq_equal} can be solved to get,
\begin{equation}\label{2eq_main2}
    \Phi(K_{b}\tau^*_{0})=\mu^*_{2}\ln{2}.
\end{equation}
Now, from \eqref{2eq_sol2}, \eqref{2eq_main1} and \eqref{2eq_main2},
\begin{equation}\label{2eq_main3}
    \frac{\gamma_{1}}{1+\frac{\gamma_{1}\tau^*_{0}}{\tau^*_{1}}}\frac{\Phi(K_{b}\tau^*_{0})}{\Phi\hspace{-0.5mm}\left(\hspace{-0.5mm}\frac{\gamma_{1}\tau^*_{0}}{\tau^*_{1}}\hspace{-0.5mm}\right)\hspace{-0.5mm}}+\frac{\Gamma_{b}}{1+K_{b}\tau^*_{0}}=\Phi(K_{b}\tau^*_{0}).
\end{equation}
As a result, using \eqref{2eq_obv} and \eqref{2eq_main3}, we have
\vspace{-1.5mm}
\begin{equation}\label{2eq_main4}
    \Phi(K_{b}\tau^*_{0})\left[1\hspace{-0.5mm}-\hspace{-0.5mm}\frac{\gamma_1}{2^{\frac{\delta}{\tau^*_{1}}}}\frac{1}{\Phi\hspace{-0.5mm}\left(2^{\frac{\delta}{\tau^*_{1}}}-1\hspace{-0.5mm}\right)}\right]=\hspace{-0.5mm}\frac{\Gamma_{b}}{1+K_{b}\tau^*_{0}}.
\end{equation}
Finally, using $K_{b}\hspace{-0.5mm}=\hspace{-0.5mm}\frac{\Gamma_{b}}{1-\left[\frac{\tau^*_{1}}{\gamma_{1}}\left(2^{\frac{\delta}{\tau^*_{1}}}-1\right)+\tau^*_{1}\right ]}$ obtained from \eqref{eq_obv}, \eqref{2eq_obv}, \eqref{2eq_equal}, in \eqref{2eq_main4}, a univariable eq. in $\tau^*_{1}$ can be written as:
\begin{eqnarray}\label{2eq_tau1}
    &\Phi\Bigl(g\left(\tau^*_{1}\right)\Bigr)\left[1\hspace{-0.5mm}-\hspace{-0.5mm}\frac{\gamma_1}{2^{\frac{\delta}{\tau^*_{1}}}}\frac{1}{\Phi\hspace{-0.5mm}\left(2^{\frac{\delta}{\tau^*_{1}}}-1\hspace{-0.5mm}\right)}\right]-\frac{\Gamma_{b}}{1+g(\tau^*_{1})}=0,
\end{eqnarray}
\noindent where $g(\tau^*_{1})=\frac{\Gamma_{b}\left[\tau^*_{1}\left(2^{\frac{\delta}{\tau^*_{1}}}-1\right)\right]}{\gamma_1\left[1-\left\{\frac{\tau^*_{1}}{\gamma_{1}}\left(2^{\frac{\delta}{\tau^*_{1}}}-1\right)+\tau^*_{1}\right \}\right]}$.
 
In order to solve \eqref{2eq_tau1} in $\tau^*_{1}=\tau_{1b}$, we use Golden Section Line Search (GSLS) method \cite{deepaktgcn37}. The number of computations $N_{C}^{GS}$ in GSLS algorithm is given by $   N_{C}^{GS}=\left \lceil 2-2.08\,{\ln\left(\frac{\xi}{\tau_{1b}^{U}-\tau_{1b}^{L}}\right)} \right \rceil$, where $\xi$ is the acceptable tolerance, $\tau_{1b}^{L}$ and $\tau_{1b}^{U}$ are respectively the lower and upper bound on $\tau_{1b}$.

\subsection{Proposed Algorithm Implementation}
Here we propose Algorithm \ref{AL1} to summarize semi-closed-form for globally optimal TA solution. The case becomes infeasible, when $R_1\triangleq R^{th}_{1}$ cannot be met even after allocating all the resources to PU. Thus, $R^{th}_{1}\hspace{-0.5mm}=\hspace{-0.5mm}\tau^{th}_{1}\log_{2}\left(1+\frac{\gamma_1\tau^{th}_{0}}{\tau^{th}_{1}}\right)$, where $\tau^{th}_{0}\triangleq \frac{f(\gamma_{1})-1}{\gamma_{1}+f(\gamma_{1})-1}$ and $\tau^{th}_{1}\triangleq \frac{\gamma_{1}}{{\gamma_{1}+f(\gamma_{1})-1}}$ from \eqref{eq_tauE} and \eqref{eq_tauI} respectively, because $N=0$, $\Gamma_a=\gamma_1$ at infeasibility. So, Algorithm \ref{AL1} starts with a value of $\delta$, and tests the feasibility conditions. It returns the feasible KKT point $(\tau^*_{k},\mu^*_{1},\mu^*_{2})$, which is the globally-optimal TA solution.

\begin{algorithm}
{\small
\caption{Semi-closed-form globally-optimal TA solution}\label{AL1}
\begin{algorithmic}[1]
\Require $\delta$, $R^{th}_1$, $\tau^L_{1b}$, $\tau^U_{1b}$ and $\zeta>0$
\Ensure $\tau^*_{k}\, \forall k\in \mathcal{N}_{0}$, $\mu^*_{1}$, $\mu^*_{2}$
\If {${\delta}>R^{th}_{1}$}\;\;this case is infeasible
\ElsIf{${\delta} < R_{1a}$, where $R_{1a}$ is the value of $R_{1}$ at $\tau_0=\tau_{0a}$ from \eqref{eq_tauE} and $\tau_1=\tau_{1a}$ from \eqref{eq_tauI}}
\State \hspace{5mm} $\tau^*_{k}=\tau_{ka}$ using \eqref{eq_tauE} and \eqref{eq_tauI}, $\mu^*_{1}=0$, $\mu^*_{2}=\mu_{2a}$
\ElsIf{$R_{1a}\le \delta\le R^{th}_{1}$}
\State \hspace{5mm} Using GSLS $\tau^*_{1}=\tau_{1b}$
\State \hspace{5mm} $\tau^*_{0}=\tau_{0b}\triangleq \frac{\tau_{1b}}{\gamma_{1}}\left(2^{\frac{\delta}{\tau_{1b}}}-1\right)$ by \eqref{2eq_obv}, $K_{b}=\frac{\Gamma_{b}}{1-\tau_{0b}-\tau_{1b}}$
\State \hspace{5mm} $\tau^*_{j}=\tau_{jb}\triangleq \frac{\gamma_{j}}{K_{b}}$ $\forall j\in \{2,3,...N+1\}$ using \eqref{2eq_equal}
\State \hspace{5mm} $\mu^*_{2}=\mu_{2b}\triangleq \frac{\Phi(K_{b}\tau_{0b})}{\ln2}$ as obtained using \eqref{2eq_main2}
\State \hspace{5mm} $\mu^*_{1}=\mu_{1b}\triangleq \frac{\mu_{2b}\ln2}{\Phi\left(\gamma_{1}\frac{\tau_{0b}}{\tau_{1b}}\right)}-1$ as obtained using \eqref{2eq_sol2}
\EndIf
\end{algorithmic}
}
\end{algorithm}

\vspace{-3mm}
\section{Performance Evaluation and Conclusion}
In this section, we first numerically validate the proposed optimal TA and then compare it against two existing TA schemes: 1) a near-optimal TA solution \cite{wcl} as found without considering any $\delta$, and 2) a uniform TA with $\tau_k=\frac{T}{N+1}, \forall k\in \mathcal{N}_{0}$. Unless explicitly stated, we have taken $L=21$ m, $N\hspace{-0.5mm}=\hspace{-0.5mm}4$, $P_{0}\hspace{-0.5mm}=\hspace{-0.5mm}10$ W,  $\sigma^{2}\hspace{-0.5mm}=\hspace{-0.5mm}-100$ dBm, $\eta=0.5$, $\delta\hspace{-0.5mm}=\hspace{-0.5mm}18$ bps/Hz, $\beta \hspace{-0.5mm}=\hspace{-0.5mm}\alpha d \hspace{-0.5mm}^{-\zeta}$, where $\alpha\hspace{-0.5mm}=\hspace{-0.5mm}\left(\frac{3\times10^{8}}{4\pi \nu}\right)^{2}$ being the average channel attenuation at unit reference distance with transmitter frequency $\nu\hspace{-0.5mm}=\hspace{-0.5mm}915$ MHz, $d$ is the distance between two TR pairs, and $\zeta=3$ is the path-loss exponent. Practically, the values of average harvested energies by PU and SU depend upon the underlying channel characteristics.

\par In Fig. \ref{valid}, for validating over optimal TA solution, we plot the system throughput against the WPT time for different sets of $N$ and $\delta$. The optimal TA values as obtained using the proposed algorithm are found to exactly match the maximum $R_{sum}$ in the feasible region. However, when $\delta$ is high, the SUs' throughput degrades due to lesser TA for SU, resulting in reduced $R_{sum}$ at optimal $\tau^*_{0}$. Further, when $N$ is large, more number of SUs leads to more harvested energy for PU as well as other SUs, hence the overall throughput increases. 

\par Fig. \ref{impact} depicts the impact of system parameters on $R_{sum}$, thereby comparing the performance of different TA schemes. In Fig. \ref{impact}(a), increase in $L$ increases the distance between two TR pairs, which leads to reduced channel gain. Further, if $\eta$ is high in Fig. \ref{impact}(b), more power can be harvested through EH, thereby enhancing $R_{sum}$. In Fig. \ref{impact}(c), $R_{sum}$ degrades after $N=4$ which is optimal, as further increase in $N$ (for a fixed $L$) curtails the TA for WPT as well as WIT. In Fig. \ref{impact}(d), benchmark approaches proposed scheme for lower values of $\delta$, but still lagging behind as it doesn’t consider all the EH possibilities. However, for large $\delta$, system performance drastically improves for proposed scheme. Similarly with respect to $L$, $\eta$ as well as $N$, the proposed scheme performs significantly better than other two. This is because, unlike in proposed scheme, the effect of a minimum PU throughput is not taken into consideration in other TA schemes, and hence the throughput cannot be maximized for higher $\delta$ corresponding to QoS-aware applications. In nutshell, on an average against all the parameters, proposed scheme is found to outperform benchmark scheme by $70\%$ and uniform fixed TA scheme by $101\%$.

So, to summarize, we proposed a novel WPT framework in a CR scenario, by exploiting all possible RF-EH opportunities. We obtained semi-closed-form for globally optimal TA solution by maximizing the system throughput. The results are numerically validated, and remarkable performance enhancement is achieved over benchmark and uniform TA schemes.

\vspace{-4mm}
\makeatletter
\renewenvironment{thebibliography}[1]{%
  \@xp\section\@xp*\@xp{\refname}%
  \normalfont\footnotesize\labelsep .5em\relax
  \renewcommand\theenumiv{\arabic{enumiv}}\let\p@enumiv\@empty
  \vspace*{-1pt}
  \list{\@biblabel{\theenumiv}}{\settowidth\labelwidth{\@biblabel{#1}}%
    \leftmargin\labelwidth \advance\leftmargin\labelsep
    \usecounter{enumiv}}%
  \sloppy \clubpenalty\@M \widowpenalty\clubpenalty
  \sfcode`\.=\@m
}{%
  \def\@noitemerr{\@latex@warning{Empty `thebibliography' environment}}%
  \endlist
}
\makeatother


\end{document}